\documentclass[twocolumn]{aa}
\usepackage{graphicx}
\usepackage{txfonts}
\begin{document}
\def\igr{IGR~J19140$+$0951}

   \title{Discovery of a new {\it INTEGRAL} source: IGR~J19140$+$0951\thanks{Based on observations with {\it INTEGRAL}, an ESA project with instruments and science data center funded by ESA and member states (especially the PI countries: Denmark, France, Germany, Italy, Switzerland, and Spain), the Czech Republic, and Poland and with the participation of Russia and the US.}}

   \author{D.C. Hannikainen
          \inst{1}
          \and
          J. Rodriguez\inst{2,3} \and
          C. Cabanac\inst{4} \and
          J. Schultz\inst{1} \and
          N. Lund\inst{5} \and
          O. Vilhu\inst{1} \and
          P.O. Petrucci\inst{4} \and
          G. Henri\inst{4}
     } 

   \offprints{D.C. Hannikainen: diana@astro.helsinki.fi}

   \institute{Observatory, PO Box 14, FIN-00014 University of
              Helsinki, Finland 
	      \and
             Centre d'Etudes de Saclay, DAPNIA/Service
              d'Astrophysique (CNRS FRE 2591), Bat. 709, Orme des
              Merisiers, Gif-sur-Yvette Cedex 91191, France 
         \and
             {\it INTEGRAL} Science Data Center, Chemin d'\'Ecogia 16,
              CH-1290 Versoix, Switzerland 
         \and
             Laboratoire d'Astrophysique, Observatoire de Grenoble, BP
              53X, 38041 Grenoble, France
         \and
             Danish Space Research Institute, Juliane Maries Vej 30, 
             2100 Copenhagen {\O}, Denmark
             }

   \date{}

   \abstract{IGR~J19140$+$0951 (formerly known as IGR~J19140$+$098) was
   discovered with the {\it INTEGRAL} satellite in March 2003.
    We report the details of the discovery, using an improved position
    for the analysis.
   We have performed a simultaneous study
   of the 5--100~keV JEM-X and ISGRI spectra from which we can
   distinguish two different states. From the results of our analysis
   we propose that IGR~J19140$+$0951 is a persistent Galactic X-ray binary,
   probably hosting a neutron star although a black hole cannot be 
   completely ruled out. 

}

   \maketitle
%

\section{Introduction}
The European Space Agency's INTErnational Gamma-Ray Astrophysical
   Laboratory ({\it INTEGRAL}) was successfully launched on 2002 Oct 17. 
The {\it INTEGRAL} payload consists of two gamma-ray instruments, 
two X-ray monitors and an optical monitor. \\
The Imager on Board the {\it INTEGRAL} spacecraft (IBIS, Ubertini et
   al. 2003) is a coded mask instrument designed for high angular
  resolution (12 arcmin, but source location down to 1 arcmin) imaging
  in the energy range from $\sim20$~keV to $\sim10$~MeV.
Its total total field of view is $29^{\circ}\times29^{\circ}$ for zero
  response with a uniform sensitivity within the central
  $\sim10^{\circ}\times10^{\circ}$.
The {\it INTEGRAL} Soft Gamma-Ray Imager (ISGRI, Lebrun et al. 2003) is the
  top layer of the IBIS detection plane, 
  and covers the energy range from 13~keV to a few hundred keV. \\
The Joint European X-ray monitor, JEM-X (Lund et al. 2003), consists 
  of two identical coded mask instruments designed for X-ray imaging in
  the range 3--35~keV with an angular resolution of 3 arcmin and a
  timing accuracy of 122~$\mu$s. 
  During our observation only the JEM~X-2 unit was being used.

 \indent Since the start of normal observations in early 
  2003, {\it INTEGRAL} has discovered a number of new transient 
  gamma- and X-ray sources.
IGR~J19140$+$0951 was discovered in the region tangent to the 
  Sagittarius spiral arm during observations targeted on
  GRS~1915$+$105 performed from 2003 March 6 through 7 
  (Hannikainen, Rodriguez \& Pottschmidt 2003).
The position of the source (Hannikainen et al. 2003) obtained 
  with an early version of the {\it Offline Scientific Analysis} 
  software (OSA) was within the error contour of a weak X-ray source 
  EXO 1912+097 (Lu et al. 1996). 
A ToO performed on IGR~J19140$+$0951 with the 
  {\it Rossi X-ray Timing Explorer} 
  allowed the absorption column density ${\mathrm{N}}_{\mathrm{H}}$ to be 
  estimated to $\sim 6\times 10^{22}$ cm$^{-2}$ 
  (Swank \& Markwardt 2003). 
Recently a (likely orbital) period of 13.55 days has been 
  obtained from re-analysis of the {\it RXTE}/{\it All Sky Monitor} 
  (Corbet, Hannikainen, Remillard 2004), suggesting a  binary 
  nature of the source. \\
\indent In this Letter we report the details of the discovery of the 
  source with {\it INTEGRAL}, study its temporal variability as well 
  as spectral evolution on timescale $\sim30$~min over this
  observation. 
In Sec.~2 we give the details of the data reduction methods that are 
  employed in the course of this analysis. 
We then present our results in Sec.~3, giving in particular the 
 most accurate position of the source (Cabanac et al. 2004), 
 and discuss our findings in the last part of the letter.


\section{Observations and data reduction}

\begin{figure}
   \centering
   \includegraphics[width=8cm]{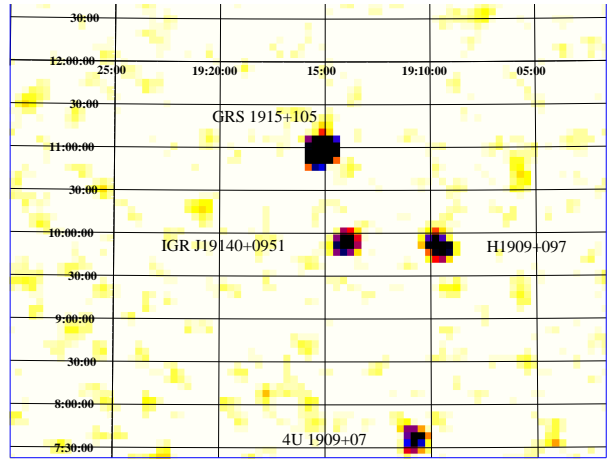}
      \caption{The IBIS/ISGRI 20--30~keV image, showing the new source
              and three other sources in the
              field-of-view. GRS~1915$+$105 is bright in this energy
              range. The image is $\sim7^{\circ}$ width and
              $\sim5^{\circ}.3$ height. North is up and East is to the
              left. The other sources in the field are dealt with in 
              Molkov et al. (2004)}
         \label{image}
   \end{figure}

The {\it INTEGRAL} observation was undertaken using the hexagonal dither
   pattern (Courvoisier et al. 2003): this consists of a hexagonal
   pattern around the nominal target location (1 source on-axis
   pointing, 6 off-source pointings, each 2 degrees apart). 
The entire duration of a pointing (science window) is 2200~s, but
   after applying a good time interval correction the effective
   exposure time is $\sim1700$~s. 
The observations were continuous, except for a short slew between each
   science window. \\
\indent The JEM X-2 data were reduced using OSA 3.0 software, following 
  the standard procedure explained in the cookbook. 
This was especially useful
  for the spectral extraction. In this case we forced the extraction of data
  products for IGR~J19140$+$0951, giving to the software the updated 
  position of the source discussed in this Letter.
The resultant spectra were grouped so that each new bin contained a 
  minimum of 60 counts, and systematic uncertainties
  (P. Kretschmar, priv. comm.) have been applied 
  as follows:
  10\% between channels 59 and 96 (4$-$7.04 keV), and 2\% above
  channel 97 ($>7.04$~keV).\\
\indent The IBIS/ISGRI data were reduced using pre-OSA 4.0 version of 
  the software. 
This new software includes the same core as OSA 3.0 except 
  updated patches for ibis\_isgr\_energy (v5.1), ibis\_isgr\_deadtime (v4.2),
  ii\_shadow\_build (v1.4), ii\_shadow\_ubc (v2.7), ii\_skyimage (v6.7.2) \& 
  ii\_spectra\_extract (v2.2), which fix many of the OSA3.0 known issues.
We made two runs of the software up to the IMA level, i.e. production
   of images.
During the first run we extracted images from individual  
 science windows in two energy ranges (20--40 keV, and 40--80 keV), 
  as well as a mosaic in the same energy range.
Figure~\ref{image} shows a zoomed IBIS/ISGRI image of the field of the new
  transient. 
The standard ISDC catalogue v13 was given as an input, and the
   software was let free to find the most significant peaks in the
   images. 
This provided us with the best position for the source which was used 
  (together with the JEM-X position) to update the entry of 
  IGR~J19140$+$0951 in the standard catalogue. 
This first run was also used  to identify the sources clearly seen 
  during our observation (only 7 were detected in the 20--40 keV
  mosaic). 
We then created a second catalogue containing only those sources. 
This second catalogue was given as the input for the second run, and 
  we forced the software to extract the source count rate in every
  science window at the position of the catalogue. 
Note that the same process was re-applied in the 20--40 keV and 40--80
   keV energy ranges, to obtain the ``true'' lightcurves of the source.
They are shown in Fig.~\ref{lc}.
%
We then extracted spectra from each science window with the Least
  Square Method.
A preliminary Crab-corrected response matrix rebinned to 16 spectral channels 
  was used in the extraction process and then in the subsequent
  fitting process.
The resultant spectra were further grouped so that each new bin had a 
  minimum of 20 counts, while 5\% systematics have been
  applied to all channels (Goldwurm et al. 2003). 
The spectra were then fitted in XSPEC v11.3, with a newly available 
  ancillary response file (P. Laurent, priv. comm.). 
We retained the energy channels between 
5 and 25 keV for JEM X-2 and those between 20 and 100 keV for ISGRI.

\section{Results}
\subsection{Refining the position of IGR~J19140$+$0951}
The source was discovered soon after the observation began 
  (Hannikainen et al. 2003) in near real time data, using an early
  version of the software (OSA1.0).
It was first spontaneously detected in science window 3 at a level of 
  $\sim3$~cts/s in the 20--40 keV (26 mCrab), and reached a level 
  of $\sim6$ cts/s ($\sim52$~mCrab) in the following science window. 
In the latter it was even detected above 40 keV, at a level of 
  $2.4$~cts/s ($\sim35$~mCrab). 
The source position had been obtained using only those science windows where the 
  source was spontaneously detected by the software in ISGRI. 
Concerning the JEM~X-2 data reduction, we used the ``JEM-X offline 
   software'' (Lund et al. 2004) to constrain with more accuracy 
   the new position. \\
We have refined the position using both JEM X-2 and ISGRI data. 
IGR~J19140$+$0951 is clearly detected in nine independent science windows of 
  the whole observing programme. 
Among them, the source was detected in two energy bands (8.4--14~keV
  and 14--35~keV) three times, 
  thus we used those 12 independent detections to derive
a best (JEM-X) weighted mean position of (J2000, errors at 1.64 $\sigma$):\\
$\mathrm{RA} =19^{\mathrm{h}} 14^{\mathrm{m}} 01^{\mathrm{s}} \pm 9s$ and 
$\mathrm{Dec} =9^\circ53^{\prime} 21^{\prime\prime} \pm 1.3^{\prime}$.\\
In the same way IGR J19140+0951 is clearly detected in IBIS/ISGRI 
  mosaics (Fig.~\ref{image}) in both energy ranges. 
We can derive a best (ISGRI) position of (J2000):\\
$\mathrm{RA} =19^{\mathrm{h}} 14^{\mathrm{m}} 02.7^{\mathrm{s}} \pm
  2^\prime$ 
  and 
  $\mathrm{Dec} =9^\circ53^{\prime} 13^{\prime\prime} \pm 2^{\prime}$ 
  (all errors are at the 90\% confidence level, see e.g. Gros et al. 2003). 
From these two independent data sets we can estimate the most accurate 
  (weighted mean) position of the source of :\\
$\mathrm{RA} =19^{\mathrm{h}} 14^{\mathrm{m}} 02^{\mathrm{s}}$ and 
$\mathrm{Dec} =9^\circ53.3^{\prime}$ (1.3$^\prime$ error at 90\%, Cabanac 
  et al. 2004). 
The source is 5.2$^{\prime}$ away from  
   EXO~1912$+$097 (Lu et al. 1996). 
As the {\it EXOSAT} error box is 6$^{\prime}$ it is possible that the
{\it EXOSAT}
detection represents an earlier outburst of the source seen by {\it INTEGRAL}.
The {\it EXOSAT} source was discovered using the demodulation technique 
  (Lu et al. 1996), but besides this detection nothing is known about
  this source.

\subsection{Temporal variability}
\begin{figure}
   \centering
    \includegraphics[width=9cm]{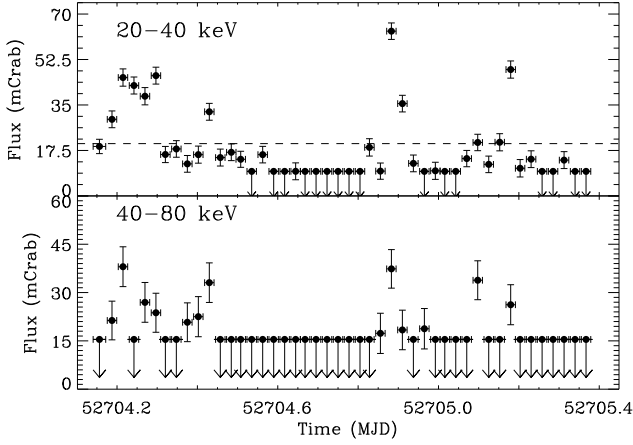}
   \caption{The 20--40~keV (top) and 40--80~keV (bottom) 
            ISGRI lightcurves. Each symbol represents a 
	    science window. The 3-sigma upper limits are denoted
            with an arrow. The dashed line in the upper panel indicates
            the 20~mCrab level.}
              \label{lc}%
    \end{figure}

Figure~\ref{lc} shows the 20--40~keV and 40--80~keV lightcurves 
  during Revolution 48.
It is immediately apparent that the source is variable on the
  timescale of 2200 seconds (typical duration of a science window) 
  during the observation. 
In the 20--40 keV range the source is detected at a flux 
  higher than the 3-$\sigma$ limit of~$9-10$ mCrab in 70$\%$ of the
  science windows. 
It is found at a level of $\sim20$~mCrab in the 20--40 keV range 50$\%$ of the 
  time, and undergoes flares on rather short timescales up to a level of 70 
  mCrab on one occasion.
The flares in the 20--40~keV range are accompanied by flaring also in
  the 40--80~keV range, reaching levels of $\sim$38~mCrab.

\subsection{Spectral analysis}

\begin{figure}
  \begin{center}
    \includegraphics[width=9cm,height=5cm]{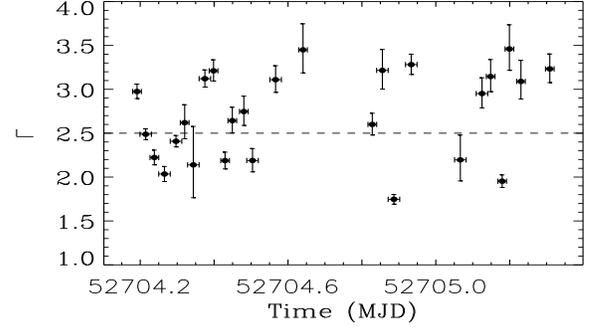}
    \caption{The powerlaw photon index for the science windows for which a good
      fit was obtained. The horizontal dashed line shows the
      $\Gamma=2.5$ level. }\label{fig:gamma}
  \end{center}
\end{figure}

\begin{figure*}
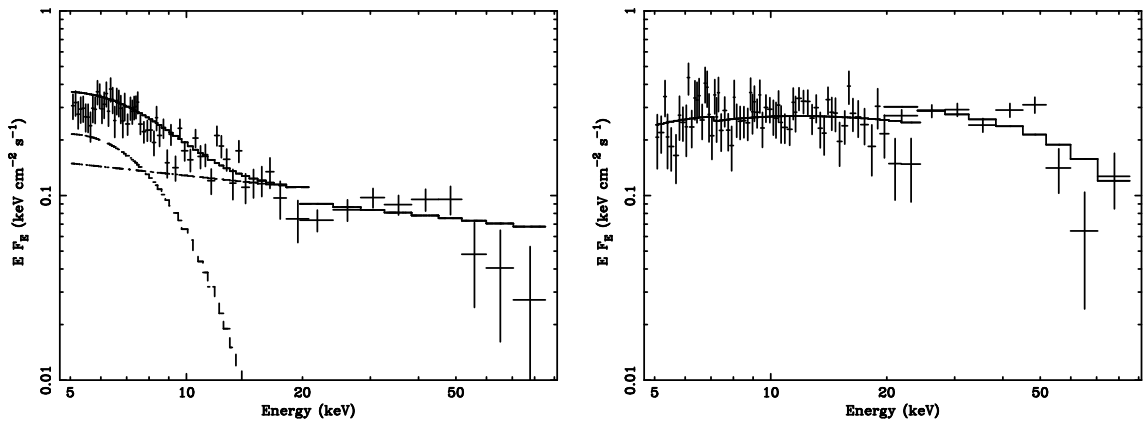

  \begin{tabular}{cc}
    \includegraphics[width=5.5cm,angle=-90]{Figure4a.eps} &
    \includegraphics[width=5.5cm,angle=-90]{Figure4b.eps} \\
  \end{tabular}    
    \caption{The faint (left) and bright (right) spectra with the
   best-fit model superimposed, a blackbody $+$ powerlaw for the faint
   spectrum and comptt for the bright spectrum.} \label{spectra}
\end{figure*}

To begin our spectral analysis, we extracted spectra from each 
   one of the 46 science windows from both JEM X-2 and ISGRI, as explained in Sec. 2.
Based on the lightcurve shown in Fig.~\ref{lc}, we selected only the
   science windows where
IGR~J19140$+$0951 is clearly detected at a significance level greater
   than $3\sigma$ in the 20--40 keV
range. We then fitted the JEM X-2 and ISGRI spectra simultaneously, with
a simple model consisting of an absorbed power law. The value of 
${\mathrm{N}}_{\mathrm{H}}$, was frozen to the value obtained with
   {\it RXTE} (Swank \& Markwardt 2003), i.e. $6\times10^{22}$~cm$^{-2}$, since the useful
energy range of JEM X-2 does not allow us to obtain a better constraint on this parameter.
We did a first run with a multiplicative constant to account for
   cross-calibration of the instruments, but it was found to be very
   close to 1 in each spectrum. 
Therefore, in a second run no such constant was included.
Fig.~\ref{fig:gamma} shows the results obtained for the
   science windows for which a good fit was achieved.
This excludes three science windows.\\
\indent To increase our statistics, we further averaged all the spectra from 
  the science windows in which IGR~J19140$+$0951 is found at a 
  flux up to $\sim$20~mCrab between 20 and 40 keV (Fig.~\ref{lc}; 
  hereafter this spectrum is referred to as ``faint'').
In addition we also averaged together all the spectra 
  where the source was found to be at a level of
  $>$20~mCrab (referred to as ``bright'').
The {\sc ftool} {\sc mathpha} was used to compute the true weighted average
   spectrum (K. Ebisawa, priv. comm.). 
Fig.~\ref{spectra} shows the spectra obtained after the averaging
   processes. 
Although a simple model fits the single spectra well, it gives a 
  relatively poor reduced chi square for the the average spectra 
  (1.55 for 65 dof in the case of the ``faint'' 
  spectrum, and 1.48 for 73 dof, in the case of the ``bright'' spectrum).

{\bf Faint spectrum.} Adding a blackbody to the 
  simple powerlaw improves the fit to a reduced $\chi^2=1.19$
  (63 dof). 
An F-test indicates that the blackbody 
  component is required at a level greater than 99.99\%.
The temperature is kT=$1.27^{+0.07}_{-0.08}$~keV and
  $\Gamma=2.39\pm0.11$.
The 2--20~keV (20--100 keV) unabsorbed flux is
  9.80$\times10^{-10}$~erg~s$^{-1}$~cm$^{-2}$
  (1.96$\times10^{-10}$~erg~s$^{-1}$~cm$^{-2}$).
Fig.~\ref{spectra} (left) shows the faint spectrum with the best-fit
  model.

{\bf Bright spectrum.} The blackbody is only marginally 
  required with an F-test probability of 92\%.
However, adding a high energy cutoff to the simple powerlaw 
  improved the fit to 
  a reduced $\chi^2=0.93$ which leads to an F-test probability 
  of $>$99.99\%.
The powerlaw photon index is 2.03$\pm0.04$.
The cutoff energy is 49$\pm3$~keV and the folding energy is
  16$^{+4}_{-7}$~keV.
Since the cutoff in a powerlaw is attributed to thermal Comptonization 
  we also fitted the bright spectrum with comptt (Fig.~\ref{spectra}, right). 
Given the energy range, the temperature of the input photons 
  was frozen to 0.5 keV.
The electron temperature is 15.1$^{+2.5}_{-1.6}$~keV and 
  the optical depth of the plasma is 2.1$^{+0.2}_{-0.3}$.
The reduced $\chi^2$ is 1.07 for 71 dof. 
The 2--20~keV (20--100 keV) unabsorbed flux is
  1.01$\times10^{-9}$~erg~s$^{-1}$~cm$^{-2}$
  (5.39$\times10^{-10}$~erg~s$^{-1}$~cm$^{-2}$).
Adding a blackbody and fixing its parameters to those of the faint
  spectrum leads to a very bad fit, ruling out a constant blackbody
  emission. 

\section{Discussion}

The refined position has allowed us to perform an improved
  analysis of IGR~J19140$+$0951 using both JEM-X and ISGRI data.
In particular, this has enabled us to obtain the true ISGRI lightcurve
  on a timescale of $\sim$2000~s as well as individual JEM~X-2 and ISGRI
  spectra. 
The ISGRI lightcurve shows that the source is variable on the
  timescale of a science window, so this would imply a maximum size of
  the emitting region of $\sim7\times10^{13}$~cm, i.e. $\sim$4 AU.
This, together with the newly-discovered period of 13.55~days, 
  implies the Galactic origin of IGR~J19140$+$0951.
It is interesting to note that throughout the 100~ksec observation, 
  the source went from being undetectable in the {\it INTEGRAL}
  instruments to a level of 80~mCrab in the 20--40~keV ISGRI range.
The variations appear to be not only related to a global change in
  luminosity but rather reflect changes in the emitting media --
  for example the appearance and possible disappearance of a blackbody 
  component in the spectra.
This is reminiscent of X-ray binaries (e.g. Tanaka \& Shibazaki 1996) and
  the newly-discovered period of 13.55~days (Corbet et al. 2004)
  strongly points to the binary nature of IGR~J19140$+$0951. \\
\indent The spectral parameters obtained for this object could be
  consistent with both types for the primary, i.e. either a neutron
  star or a black hole. 
In fact, although neutron stars usually have a lower energy cutoff in
  their spectra, some black holes can show a cutoff as
  low as 30~keV (e.g. XTE~J1550$-$564, Rodriguez et al. 2003).
However, in the latter the low energy of the cutoff is accompanied by
  the very bright emission of soft X-rays (close to 1 Crab in the
  1--10~keV range) which is not the case here. 
In addition, the main difference between a neutron star and a black
  hole in thermal Comptonization is related to the temperature of the
  electrons (Barret 2001). 
In the first phenomenological model we used, it is usually admitted
  that it is more the folding energy which is close to the electron
  temperature rather than the cutoff energy.
In that case, IGR~J19140$+$0951 manifests the expected difference 
  for a neutron star compared to a black hole such as XTE~J1550$-$564.
This and the persistence of the source would point to a neutron star
  rather than a black hole. 
However, a black hole cannot be dismissed since the 
  variations of the photon index (Fig.~3) are similar
  to those seen in GRS~1915$+$105 (e.g. Markwardt et al. 1999). \\
\indent The high energy tail would represent the Comptonization of
  the soft photons on relativistic electrons. 
And indeed, the averaged bright spectrum is well fitted with a thermal
  Comptonization model. 
In addition to a variation in the blackbody, or thermal, component,
  the variations may also indicate transitions between thermal
  Comptonization and non-thermal or hybrid thermal-non-thermal
  Comptonization. 
The quality of our data does not allow us to answer more precisely these
  points; a longer accumulation of data in time is currently underway
  with the aim to increase the statistics at especially the higher
  energies which in turn will allow us to address this question and
  the true nature of the compact object. \\
\indent Further analysis of this source will be deferred to a later paper
  which will include the remaining {\it INTEGRAL} observations from both
  the Open Time programme and the Galactic Plane Scans of the Core
  Programme, plus multiwavelength coverage including e.g. the Nordic
  Optical Telescope and the VLA. 

\begin{acknowledgements}
DCH gratefully acknowledges a Fellowship from the Academy of Finland. 
JR acknowledges financial support from the French space agency (CNES). 
JS acknowledges the financial support of Vilho, Yrj\"o and
Kalle V\"ais\"al\"a foundation. OV \& JS are grateful to  
the Finnish space research programme Antares and TEKES. 
The authors wish to thank Ken Ebisawa for useful suggestions,
and Aleksandra Gros and Marion Cadolle Bel for providing us 
with the most recent IBIS products. The authors also wish to thank the
referee for useful comments. 
\end{acknowledgements}

\end{document}